\documentclass[11pt]{article}

\usepackage{amstext}
\usepackage{enumerate}
\usepackage{citesort}
\usepackage[mathscr]{eucal}
\usepackage{epsfig}
\usepackage{amsmath}
\usepackage{theorem}

\def\myendproof{{\ \vbox{\hrule\hbox{%
   \vrule height1.3ex\hskip0.8ex\vrule}\hrule }}\par}
 \newtheorem{theorem}{Theorem}[section]
\newtheorem{lemma}[theorem]{Lemma}

\newtheorem{fact}[theorem]{Fact}
\newtheorem{definition}{Definition}
\newenvironment{proof}{{\it Proof. }}{\myendproof}

\newcommand{\comment}[1]{}

\title{Predicting RNA Secondary Structures with Arbitrary Pseudoknots
by Maximizing the Number of Stacking Pairs}

\author{\hspace*{.5in}
Samuel Ieong\thanks{Department of Computer Science,
Yale University, New Haven, CT 06520.}
\and
Ming-Yang Kao\thanks{Department of Computer Science,
Northwestern University, Evanston, IL 60201 (kao@cs.northwestern.edu).
This research was supported in part by NSF Grant EIA-0112934.}
\and 
Tak-Wah Lam\thanks{Department of Computer Science,
The University of Hong Kong, Hong Kong
(\{twlam, smyiu\}@cs.hku.hk).
This research was supported in part by Hong Kong RGC grant HKU-7027/98E.}
\hspace*{.5in}
\and
Wing-Kin Sung\thanks{Department of Computer Science,
National University of Singapore, 3 Science Drive 2,
Singapore 117543 (ksung@comp.nus.edu.sg).}
\and
Siu-Ming Yiu\footnotemark[3]
}

\begin{document}
\date{}
\maketitle

\begin{abstract}

The paper investigates the computational problem of predicting
RNA secondary structures.
The general belief is that allowing pseudoknots makes the problem hard.
Existing polynomial-time algorithms are heuristic algorithms
with no performance guarantee and can only handle limited
types of pseudoknots.
In this paper we initiate the study of
predicting RNA secondary structures with
a maximum number of stacking pairs while allowing arbitrary pseudoknots.
We obtain two approximation algorithms with worst-case approximation ratios 
of $1/2$ and $1/3$ for planar and general secondary structures,
respectively. For an RNA sequence of $n$ bases,
the approximation algorithm for planar secondary structures
runs in
$O(n^3)$ time while that for the general case runs in linear time.
Furthermore, we prove that allowing
pseudoknots makes it NP-hard to maximize 
the number of stacking pairs in a planar secondary structure.
This result is in contrast with the recent 
NP-hard results on psuedoknots
which are based on optimizing some general and complicated
energy functions.

\end{abstract}

\section{Introduction}
\begin{figure}[t]
\begin{centering}
\epsfig{file=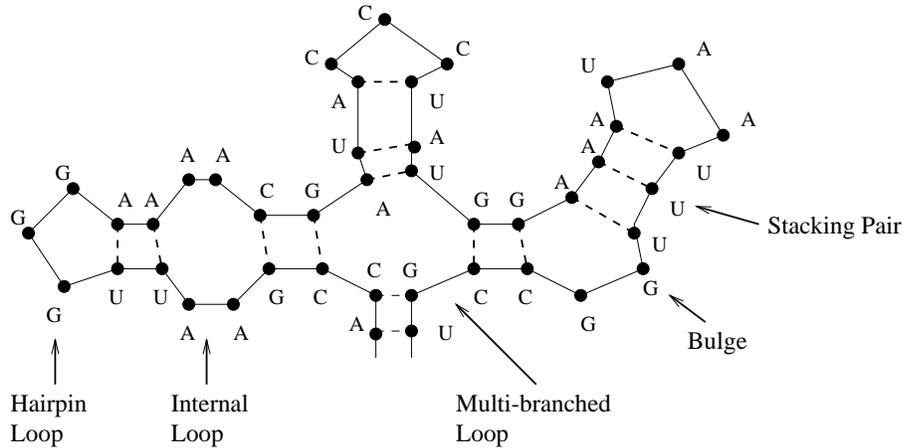, height=2.3in}
\caption{Example of a folded RNA}
\label{fig:RNA-structure}
\end{centering}
\end{figure}

Ribonucleic acids (RNAs) are molecules that are responsible for regulating many genetic 
and metabolic activities in cells.  
An RNA is single-stranded and 
can be considered as a sequence of nucleotides (also known as bases).  There are four 
basic nucleotides, namely, Adenine (A), Cytosine (C), Guanine (G), and Uracil (U).  An 
RNA folds into a 3-dimensional structure by forming pairs of bases.  Paired bases tend to 
stabilize the RNA (i.e., have negative free energy).  Yet base pairing does not occur 
arbitrarily.  In particular, A-U and C-G form stable pairs and
are known as the {\em Watson-Crick}
base pairs. Other base pairings are less stable and often ignored.  An example of a 
folded RNA is shown in Figure~\ref{fig:RNA-structure}.  
Note that this figure is just schematic; 
in practice, RNAs are 3-dimensional molecules.  

The 3-dimensional structure is related to the function of the RNA. 
Yet existing experimental techniques for determining the
3-dimensional structures of RNAs are
often very costly and time consuming (see, e.g., \cite{Meidanis:1997:ICM}).
The secondary structure of an RNA is the set of base pairings formed
in its 3-dimensional structure.
To determine the 3-dimensional structure of a given RNA sequence,
it is useful to determine the corresponding secondary structure.
As a result, it is important to design efficient algorithms to
predict the secondary structure with computers.

\begin{figure}[t]
\begin{centering}
\epsfig{file=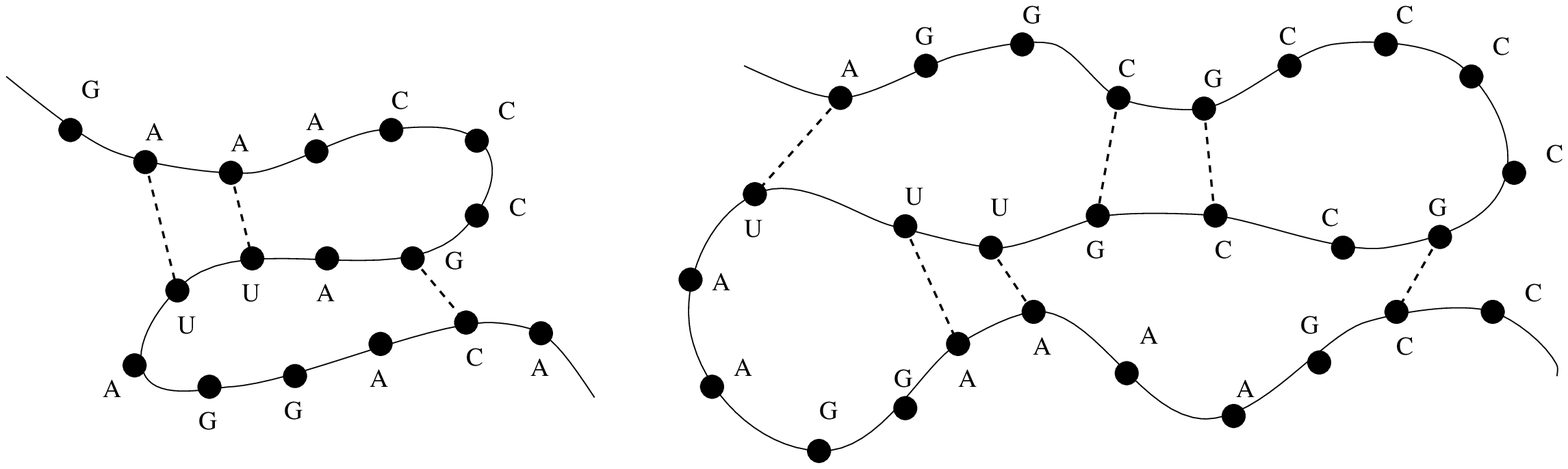, height=1.4in}
\caption{Examples of pseudoknots}
\label{fig:simple-pk}
\end{centering}
\end{figure}

{From} a computational viewpoint, the challenge of the RNA secondary
structure prediction
problem arises from some special structures called pseudoknots, 
which are defined as follows.  
Let $S$ be an RNA sequence $s_1, s_2, \cdots, s_n$.   A 
{\it pseudoknot} is composed of two interleaving base
pairs, i.e., $(s_i, s_j)$ and $(s_k, s_\ell)$
such that $i < k < j < \ell$.  See Figure~\ref{fig:simple-pk} for examples.

If we assume that the secondary structure of an RNA contains no pseudoknots, 
the secondary structure can be decomposed into a few types of loops: stacking pairs, 
hairpins, bulges, internal loops, and multiple loops
(see, e.g., Tompa's lecture notes \cite{Tompa:2000:LNB} or
Waterman's book \cite{Waterman:1995:ICB}).   A
{\it stacking pair} is a loop formed by two pairs of consecutive
bases $(s_i, s_j)$ and $(s_{i+1}, s_{j-1})$ with $i+4 \leq j$.
See Figure~\ref{fig:RNA-structure} for an example.
By definition, a stacking pair contains no unpaired bases and any other kinds 
of loops contain one or more unpaired bases.  Since
unpaired bases are destabilizing and have positive free energy, 
stacking pairs are the only type of 
loops that have negative free energy and stabilize the secondary structure.   
It is also natural to 
assume that the free energies of loops are independent.  
Then an optimal pseudoknot-free secondary structure can be 
computed using dynamic programming in $O(n^3)$ time 
\cite{Lyngso:1999:FEI,Lyngso:1999:ILR,Zuker:1984:RSS,Zuker:1989:UDP}.

However, pseudoknots are known to exist in some RNAs.  
For predicting secondary structures with pseudoknots,
Nussinov et al.~\cite{Nussinov:1978:ALM} have studied the case where
the energy function is minimized when the number of base pairs is 
maximized and have obtained an $O(n^3)$-time algorithm
for predicting secondary structures.  
Based on some special energy functions, Lyngso and Pedersen 
\cite{Lyngso:2000:RPP} have proven that determining the optimal secondary structure
possibly with
pseudoknots is NP-hard. Akutsu \cite{Akutsu:2000:DPA}
has shown that it is NP-hard to determine an optimal
planar secondary structure, where a secondary structure is {\it planar}
if the
graph formed by the base pairings and the backbone connections of adjacent bases is 
planar (see Section 2 for a more detailed definition).
Rivas and Eddy \cite{Rivas:1999:DPA}, Uemura et al. \cite{Uemura:1999:TAG},
and
Akutsu \cite{Akutsu:2000:DPA} have also proposed polynomial-time algorithms that can 
handle limited types of pseudoknots; note that the exact types of
such pseudoknots are implicit in these algorithms and difficult to
determine.

Although it might be desirable to have a better classification of pseudoknots and 
better algorithms that
can handle a wider class of pseudoknots, 
this paper approaches the problem in a different general direction.
We initiate the study of predicting RNA secondary 
structures that allow arbitrary pseudoknots while maximizing
the number of stacking pairs. 
Such a simple energy function is meaningful as
stacking pairs are the only loops that stabilize
secondary structures.  We obtain two approximation algorithms with worst-case ratios of 
1/2 and 1/3 for planar and general secondary structures, respectively.
The planar
approximation algorithm makes use of a geometric observation 
that allows us to 
visualize the planarity of stacking pairs on a rectangular grid;
interestingly, such an
observation does not hold if our aim is to maximize the number of base pairs.
This algorithm runs in $O(n^3)$ time.
The second approximation algorithm is more complicated and
is based on a combination of multiple  ``greedy'' strategies.
A straightforward analysis cannot lead to the approximation ratio of $1/3$.
We make use of amortization over different steps to obtain the desired
ratio. This algorithm runs in $O(n)$ time.  

To complement these two algorithms, we also prove
that allowing pseudoknots makes it NP-hard to 
find the planar secondary structure with the
largest number of stacking pairs.  
The proof makes use of a reduction from a 
well-known NP-complete problem called Tripartite Matching
\cite{Garey:1979:CIG}.
This result indicates that the hardness of the RNA secondary
structure prediction problem may be inherent in the pseudoknot structures 
and may not be necessarily due to the complication of the energy functions.
This is in contrast to the other NP-hardness results discussed
earlier.

The rest of this paper is organized into four sections.  Section 2
discusses some basic properties.  
Sections 3 and 4 present the approximation algorithms for
planar and general secondary structures, respectively.  
Section 5 details the NP-hardness result.  
Section 6 concludes the paper with open problems.

\section{Preliminaries}

Let $S=s_1 s_2 \cdots s_n$ be an RNA sequence of $n$ bases.
A {\it secondary structure} ${\cal P}$ of $S$ is a set of
Watson-Crick pairs $(s_{i_1}, s_{j_1}),\ldots, (s_{i_p},
s_{j_p})$, where $s_{i_r}+2 \leq s_{j_r}$ for all $r=1, \ldots, p$
and no two pairs share a base.
We denote $q$ ($q \geq 1$)
consecutive stacking pairs ($s_i, s_j$), ($s_{i+1}, s_{j-1}$);
($s_{i+1}, s_{j-1}$), ($s_{i+2}, s_{j-2}$)
$\ldots$ ($s_{i+q-1}, s_{j-q+1}$), ($s_{i+q}, s_{j-q}$) of
${\cal P}$ by ($s_i,s_{i+1}, \ldots, s_{i+q};$
\linebreak[4] $s_{j-q}, \ldots, s_{j-1}, s_j$).

\begin{definition}
Given a secondary structure ${\cal P}$,
we define an undirected
graph $G({\cal P})$ such that the bases
of $S$ are the nodes of $G({\cal P})$ and $(s_i, s_j)$ is
an edge of $G({\cal P})$ if $j = i+1$ or $(s_i, s_j)$ is
a base pair in ${\cal P}$.
\end{definition}

\begin{definition}
A secondary structure ${\cal P}$ is planar if $G({\cal P})$ is
a planar graph.
\end{definition}

\begin{definition}
A secondary structure ${\cal P}$ is said to contain an
{\it interleaving block} if ${\cal P}$ contains three 
stacking pairs
$(s_i, s_{i+1}; s_{j-1}, s_j)$, $(s_{i'}, s_{i'+1}; s_{j'-1}, s_{j'})$,
$(s_{i''}, s_{i''+1}; s_{j''-1},s_{j''})$ where $i < i' < i'' < j < j' < j''$.
\end{definition}

\begin{lemma}
\label{interleavingblock}
If a secondary structure ${\cal P}$ contains
an interleaving block, ${\cal P}$ is non-planar.
\end{lemma}

\begin{figure*}[hbtp]
\begin{center}
\scalebox{0.5}[0.5]{\includegraphics{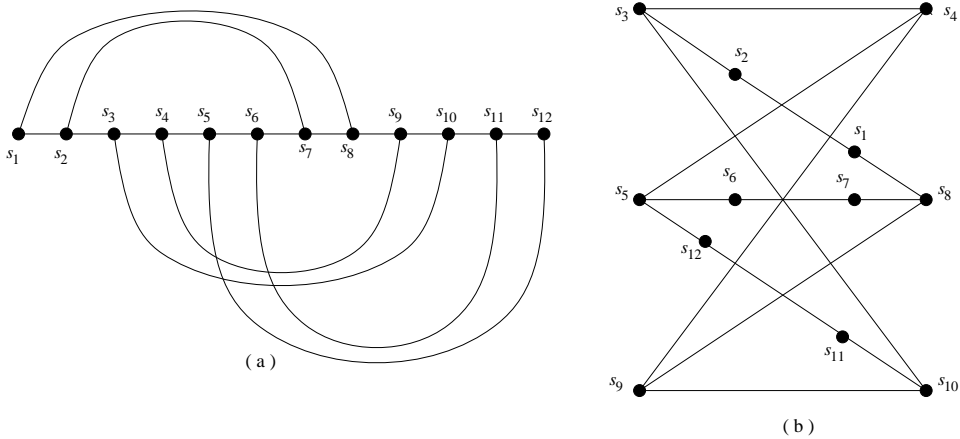}}
\caption{Interleaving block}
\label{interblock}
\end{center}
\end{figure*}

\begin{proof}
Suppose ${\cal P}$ contains an interleaving block.  Without
loss of generality, we assume that ${\cal P}$ contains
the stacking pairs ($s_1, s_2; s_7, s_8$),
($s_3, s_4; s_9, s_{10}$), and ($s_5, s_6; s_{11}, s_{12}$).
Figure \ref{interblock}(a) shows the
subgraph of $G({\cal P})$ corresponding to these
stacking pairs. Since this subgraph contains a homeomorphic copy of
$K_{3,3}$ (see Figure \ref{interblock}(b)),
$G({\cal P})$ and ${\cal P}$ are non-planar.
\end{proof}

\section{An Approximation Algorithm for Planar Secondary Structures}
We present an algorithm which,
given an RNA sequence $S = s_1 s_2 \ldots s_n$,
constructs a {\it planar} secondary structure of $S$ 
to approximate one with the maximum number of stacking pairs
with a ratio of at least $1/2$. This
approximation algorithm is based on the subtle
observation in Lemma \ref{planarembedding}
that if a secondary structure ${\cal P}$ is planar,
the subgraph of $G({\cal P})$ which contains {\it only} the stacking pairs
of ${\cal P}$ can be embedded in a grid with a useful property.
This property enables us to consider only the secondary structure of
$S$ {\it without pseudoknots} in order to achieve 1/2 approximation
ratio. 

\begin{definition}
Given a secondary structure ${\cal P}$, we define a
{\it stacking pair embedding} of
${\cal P}$ on a grid as follows.
Represent the bases of $S$ as $n$ consecutive grid points on the
same horizontal grid line $L$ such that $s_i$ and $s_{i+1}$
$(1 \leq i < n)$ are connected directly by a horizontal grid edge.
If $(s_i, s_{i+1}; s_{j-1}, s_j)$ is a stacking pair of ${\cal P}$,
$s_i$ and $s_{i+1}$ are connected to $s_j$ and $s_{j-1}$ respectively
by a sequence of grid edges such that the two sequences must
be either both above or both below $L$.
\end{definition}

\begin{figure*}[hbtp]
\begin{center}
\scalebox{0.5}[0.5]{\includegraphics{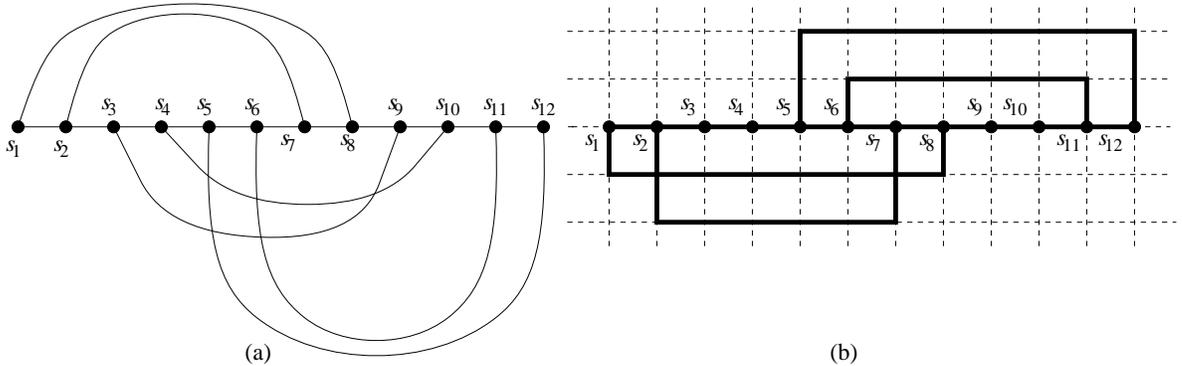}}
\caption{An example of a stacking pair embedding}
\label{embedding-eg}
\end{center}
\end{figure*}

Figure \ref{embedding-eg} shows a stacking pair embedding
(Figure \ref{embedding-eg}(b))
of a given secondary structure (Figure \ref{embedding-eg}(a)).
Note that ($s_3,s_9$)
do not form a stacking pair with other base pair, so $s_3$
is not connected to $s_9$ in the stacking pair embedding.
Similarly, $s_4$ is not connected to $s_{10}$ in the
embedding.

\begin{definition}
A stacking pair embedding is said to be {\it planar} if
it can be drawn in such a way that
no lines cross or overlap with each other in the grid.
\end{definition}

The embedding shown in Figure \ref{embedding-eg}(b) is planar.

\begin{lemma}
\label{planarembedding}
Let ${\cal P}$ be a secondary structure of an RNA sequence $S$.
Let $E$ be a stacking pair embedding of ${\cal P}$.
If ${\cal P}$ is planar, then $E$ must be planar.
\end{lemma}

\begin{proof}
If ${\cal P}$ does not have a planar stacking
pair embedding, we claim that ${\cal P}$ contains an
interleaving block. Let $L$ be the horizontal grid line
that contains the bases of $S$ in $E$.
Since ${\cal P}$ does not have a planar
stacking pair embedding, we can assume that $E$ has
two stacking pairs intersect
above $L$ (see Figure \ref{non-planar-sec-struct}(a)).

\begin{figure*}[hbtp]
\begin{center}
\scalebox{0.5}[0.5]{\includegraphics{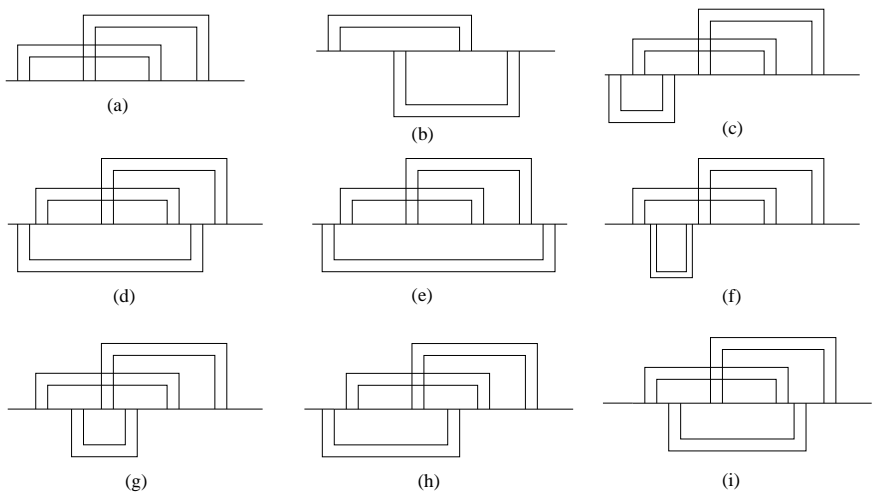}}
\caption{Non-planar stacking pair embedding}
\label{non-planar-sec-struct}
\end{center}
\end{figure*}

If there is no other stacking pair underneath these two
pairs, we can flip one of the pairs below $L$ as shown
in Figure \ref{non-planar-sec-struct}(b). So, there must be
at least one stacking pair underneath these two
pairs. By checking all
possible cases (all non-symmetric cases are shown in
Figures \ref{non-planar-sec-struct}(c) to (i)), it can be
shown that $E$ cannot be redrawn without crossing or overlapping
lines only if it contains an interleaving block
(Figures \ref{non-planar-sec-struct}(h) and (i)). So, by
Lemma \ref{interleavingblock}, ${\cal P}$ is non-planar.
\end{proof}

By Lemma \ref{planarembedding},
we can relate two secondary structures having the maximum
number of stacking pairs with and without pseudoknots
in the following lemma.

\begin{lemma}
\label{1/2-ratio}
Given an RNA sequence $S$, let $N^*$ be the maximum number of
stacking pairs that can be formed by a planar secondary
structure of $S$ and let $W$ be the maximum
number of stacking pairs that can be formed by $S$ without
pseudoknots. Then, $W \geq \frac{N^*}{2}$.
\end{lemma}

\begin{proof}
Let ${\cal P}^*$ be a planar secondary structure of $S$ with $N^*$
stacking pairs. Since ${\cal P}^*$ is planar, by Lemma
\ref{planarembedding}, any stacking pair embedding of ${\cal P}^*$
is planar.

Let $E$ be a stacking pair embedding of ${\cal P}^*$
such that no lines cross each other in the grid.
Let $L$ be the horizontal grid line of $E$ which
contains all bases of $S$.
Let $n_1$ and $n_2$ be the number of stacking pairs which
are drawn above and below $L$, respectively.
Without loss of generality,
assume that $n_1 \geq n_2$. Now, we construct another planar
secondary structure ${\cal P}$ from $E$ by deleting all stacking
pairs which are drawn below $L$.
Obviously, ${\cal P}$ is a planar secondary structure of $S$ without
pseudoknots. Since $n_1 \geq n_2$, $n_1 \geq \frac{N^*}{2}$.
As $W \geq n_1$, $W \geq \frac{N^*}{2}$.
\end{proof}

Based on Lemma \ref{1/2-ratio}, we now present the dynamic programming
algorithm $MaxSP$ which computes the maximium number of
stacking pairs that can be formed by an RNA
sequence $S=s_1 s_2 \ldots s_n$ without pseudoknots.

\vspace{5pt}
\noindent
{\bf Algorithm $MaxSP$}

Define $V(i,j)$ (for $j \geq i$) as the maximum number of stacking
pairs without pseudoknots that can be formed by $s_i \ldots s_j$
{\it if $s_i$ and $s_j$ form a Watson-Crick pair}.
Let $W(i,j)$ ($j \geq i$) be the maximum number
of stacking pairs without pseudoknots that can be formed by
$s_i \ldots s_j$. Obviously, $W(1,n)$ gives the maximum
number of stacking pairs that can be formed by $S$ without
pseudoknots.

\noindent \fbox{Basis:}

For $j = i, i+1, i+2 \mbox{~or~} i+3$ ($j \leq n$),
\[
\begin{array}{lll}
V(i,j) & = 0  & \mbox{ if $s_i, s_{j}$ form a Watson-Crick pair;} \\
W(i,j) & = 0.  &
\end{array}
\]

\noindent \fbox{Recurrence:}

For $j > i+3$,
\[
\begin{array}{llll}
W(i,j) & = & \max \left\{
\begin{array}{ll}
V(i,j) & \mbox{ if $s_i$, $s_j$ form a Watson-Crick pair} \\
W(i+1, j) & \\
W(i, j-1) &
\end{array}
\right\}; \\
&&\\
V(i,j) & = & \max \left\{
\begin{array}{l}
V(i+1, j-1) + 1 \mbox{~~~~if $s_{i+1}$, $s_{j-1}$ form a Watson-Crick pair} \\
\max_{i+1 \leq k \leq j-2}{\{W(i+1,k)+W(k+1,j-1)\}}
\end{array}
\right\}.
\end{array}
\]

\begin{lemma}
Given an RNA sequence $S$ of length $n$, Algorithm $MaxSP$
computes the maximum number of stacking pairs that can be
formed by $S$ without pseudoknots in $O(n^3)$ time and
$O(n^2)$ space.
\end{lemma}

\begin{proof}
There are $O(n^2)$ entries $V(i,j)$ and $W(i,j)$ to be
filled. To fill an entry of $V(i,j)$, we check
at most $O(n)$ values. To fill an entry of $W(i,j)$, $O(1)$ time
suffices. The total time complexity for filling all entries
is $O(n^3)$. Storing all entries requires $O(n^2)$ space.
\end{proof}

Although Algorithm $MaxSP$ presented in the above only
computes the number of stacking pairs, it can be easily modified
to compute the secondary structure.
Thus we have the following theorem.

\begin{theorem}
The Algorithm $MaxSP$ is an $(1/2)$-approximation algorithm
for the problem of constructing a secondary structure which
maximizes the number of stacking pairs for an RNA sequence $S$.
\end{theorem}

\section{An Approximation Algorithm for General Secondary Structures}
We present Algorithm $GreedySP()$ which,
given an RNA sequence $S = s_1 s_2 \ldots s_n$,
constructs a secondary structure of $S$ (not necessarily planar) 
with at least $1/3$ of the maximum possible number of
stacking pairs.
The approximation algorithm uses a greedy approach.
Figure \ref{1/3-approx-alg} shows
the algorithm $GreedySP()$.

\begin{figure}[htbp]
\fbox{
\begin{minipage}{.95\textwidth}
\noindent // Let $S=s_1 s_2 \ldots s_n$ be the input RNA sequence.
Initially, all $s_j$ are unmarked.

\noindent // Let $E$ be the set of base pairs output by the algorithm.
Initially, $E = \emptyset$.

\vspace{5pt}
\noindent $GreedySP(S, i)$
\hspace{10pt}
// $i \geq 3$

\begin{enumerate}
\item Repeatedly find the {\it leftmost} $i$ consecutive stacking pairs
      $SP$ (i.e., find $(s_p,\ldots,s_{p+i};s_{q-i},\ldots,s_q)$ such that
      $p$ is as small as possible) formed by unmarked bases.
      Add $SP$ to $E$ and mark all these bases.
\item For $k = i-1$ downto $2$, \\
      Repeatedly find {\it any} $k$ consecutive stacking pairs $SP$
      formed by unmarked bases.
      Add $SP$ to $E$ and mark all these bases.
\item Repeatedly find the {\it leftmost} stacking pair $SP$ formed
      by unmarked bases.
      Add $SP$ to $E$ and mark all these bases.
\end{enumerate}
\end{minipage}
} 
\caption{A 1/3-Approximation Algorithm}
\label{1/3-approx-alg}
\end{figure}

In the following, we analyze the approximation ratio of
this algorithm.
The algorithm $GreedySP(S, i)$ will generate a sequence of $SP$'s
denoted by $SP_1, SP_2, \ldots, SP_h$.

\begin{fact}
\label{spdisjoint}
For any $SP_j$ and $SP_k$ $(j \neq k)$, the 
stacking pairs in $SP_j$ do not share any base with those
in $SP_k$.
\end{fact}

For each $SP_j = (s_p, \ldots, s_{p+t}; s_{q-t}, \ldots, s_q)$,
we define two intervals of indexes, ${\cal I}_j$ and
${\cal J}_j$, as $[p .. p+t]$
and $[q-t .. q]$, respectively.
In order to compare
the number of stacking pairs formed with that in the optimal
case, we have the following definition.


\begin{definition}
\label{xpi}
Let ${\cal P}$ be an optimal secondary structure of $S$ with
the maximum number of stacking pairs. Let
${\cal F}$ be the set of all stacking pairs of ${\cal P}$.
For each $SP_j$ computed by
$GreedySP(S,i)$ and $\beta = {\cal I}_j$ or ${\cal J}_j$,
\[\mbox{let~}{\cal X}_{\beta} = \{ (s_k, s_{k+1}; s_{w-1}, s_w) \in
{\cal F} | \mbox{~at least one of indexes~} k, k+1, w-1, w
\mbox{~is in~} \beta\}. \]
\end{definition}

Note that ${\cal X}_\beta$'s may not be disjoint.

\begin{lemma}
\label{complete}
$\bigcup_{1 \leq j \leq h} \{{\cal X}_{{\cal I}_j} \cup
{\cal X}_{{\cal J}_j}\} = {\cal F}$.
\end{lemma}

\begin{proof}
We prove this lemma by contradiction. Suppose that there exists a
stacking pair ($s_k,s_{k+1};s_{w-1},s_w$) in ${\cal F}$ but not in
any of ${\cal X}_{{\cal I}_j}$ and ${\cal X}_{{\cal J}_j}$.
By Definition \ref{xpi}, none of the indexes, $k,k+1,w-1,w$
is in any of ${\cal I}_j$ and ${\cal J}_j$. This contradicts
with Step 3 of Algorithm $GreedySP(S,i)$.
\end{proof}

\begin{definition}
\label{x'pi}
For each ${\cal X}_{{\cal I}_j}$,
\[
\mbox{let~} {\cal X}'_{{\cal I}_j} =
{\cal X}_{{\cal I}_j} -
\bigcup_{k<j} \{ {\cal X}_{{\cal I}_k} \cup
                 {\cal X}_{{\cal J}_k} \}, 
\mbox{~and let~} {\cal X}'_{{\cal J}_j} =
{\cal X}_{{\cal J}_j} -
\bigcup_{k<j} \{ {\cal X}_{{\cal I}_k} \cup
                 {\cal X}_{{\cal J}_k} \} - {\cal X}_{{\cal I}_j} \]
\end{definition}

Let $|SP_j|$ be the number of stacking pairs represented by
$SP_j$. Let $|{\cal I}_j|$ and $|{\cal J}_j|$ be the numbers
of indexes in the intervals ${\cal I}_j$ and ${\cal J}_j$,
respectively.

\begin{lemma}
\label{sumfraction}
Let $N$ be the number of stacking pairs computed by 
Algorithm $GreedySP(S,i)$ and $N^*$ be the maximum number of
stacking pairs that can be formed by $S$.
If for all $j$, we have
$|SP_j| \geq \frac{1}{r} \times
|({\cal X}'_{{\cal I}_j} \cup {\cal X}'_{{\cal J}_j})|$, then
$N \geq \frac{1}{r} \times N^*$.
\end{lemma}

\begin{proof}
By Definition \ref{x'pi}, 
$\bigcup_k \{{\cal X}_{{\cal I}_k} \cup {\cal X}_{{\cal J}_k}\} =
 \bigcup_k \{{\cal X}'_{{\cal I}_k} \cup {\cal X}'_{{\cal J}_k}\}$.
Then by Fact \ref{spdisjoint}, $N = \sum_j |SP_j|$. Thus,
$N \geq \frac{1}{r} \times
|\bigcup_k \{{\cal X}_{{\cal I}_k} \cup {\cal X}_{{\cal J}_k}\}|$.
By Lemma \ref{complete}, $N \geq \frac{1}{r} \times N^*$.
\end{proof}

\begin{lemma}
\label{boundforspi}
For each $SP_j$ computed by $GreedySP(S,i)$, we have
$|SP_j| \geq \frac{1}{3}
\times
|({\cal X}'_{{\cal I}_j} \cup {\cal X}'_{{\cal J}_j})|$.
\end{lemma}

\begin{proof} 
There are three cases as follows.

\vspace{5pt}
\noindent
{\it Case 1:} $SP_j$ is computed by $GreedySP(S, i)$ in Step 1.
Note that $SP_j = (s_p, \ldots, s_{p+i};$ $s_{q-i}, \ldots, s_q)$ is
the leftmost $i$ consecutive stacking pairs, i.e., $p$ is the
smallest possible.
By definition, $|{\cal X}'_{{\cal I}_j}|, |{\cal X}'_{{\cal J}_j}| \leq i+2$.
We further claim that $|{\cal X}'_{{\cal I}_j}| \leq i+1$.
Then $|SP_j| / | {\cal X}'_{{\cal I}_j} \cup {\cal X}'_{{\cal J}_j}| 
\geq i/((i+1)+(i+2)) \geq 1/3$ (as $i \geq 3$).

We prove the claim by contradiction. Assume that
$|{\cal X}'_{{\cal I}_j}| = i+2$. That is,
for some integer $t$, ${\cal F}$ has $i+2$ consecutive stacking pairs
$(s_{p-1}, \ldots, s_{p+i+1}; s_{t-i-1}, \ldots, s_{t+1})$.
Furthermore, none of the bases $s_{p-1}, \ldots, s_{p+i+1}, s_{t-i-1}, \ldots, s_{t+1}$
are marked before $SP_j$ is chosen; otherwise,
suppose one such base, says $s_a$, is marked 
when the algorithm chooses $SP_\ell$ for $\ell < j$,
then an stacking pair adjacent to $s_a$ does not belong to
${\cal X}'_{{\cal I}_j}$ and they belong to ${\cal X}'_{{\cal I}_\ell}$
or ${\cal X}'_{{\cal J}_\ell}$ instead.
Therefore, $(s_{p-1}, \ldots, s_{p+i-1}; s_{t-i+1}, \ldots, s_{t+1})$
is the leftmost $i$ consecutive stacking pairs formed by unmarked bases
before $SP_j$ is chosen.
As $SP_j$ is not the leftmost $i$ consecutive stacking pairs,
this contradicts the selection criteria of $SP_j$.
The claim follows.

\vspace{5pt}
\noindent
{\it Case 2:} $SP_j$ is computed by $GreedySP(S, i)$ in Step 2.
Let $|SP_j| = k \geq 2$. Let 
$SP_j = (s_p, \ldots, s_{p+k}; s_{q-k}, \ldots, s_q)$.
By definition, $|{\cal X}'_{{\cal I}_j}|, |{\cal X}'_{{\cal J}_j}| \leq k+2$.
We claim that $|{\cal X}'_{{\cal I}_j}|, |{\cal X}'_{{\cal J}_j}| \leq k+1$.
Then $|SP_j| / | {\cal X}'_{{\cal I}_j} \cup {\cal X}'_{{\cal J}_j}|
\geq k/((k+1)+(k+1))$,
which is at least $1/3$ as $k \geq 2$.

To show that $|{\cal X}'_{{\cal I}_j}| \leq k+1$ by contradiction,
assume $|{\cal X}'_{{\cal I}_j}| = k+2$. Thus, for some integer $t$,
there exist $k+2$ consecutive stacking pairs
$(s_{p-1}, \ldots, s_{p+k+1}; s_{t-k-1}, \ldots, s_{t+1})$.
Similarly to case 1, we can show that
none of the bases $s_{p-1}, \ldots, s_{p+k+1}, s_{t-k-1}, \ldots, s_{t+1}$
are marked before $SP_j$ is chosen.
Thus, $GreedySP(S, i)$ should select some $k+1$ or $k+2$ consecutive
stacking pairs
instead of the chosen $k$ consecutive stacking pairs,
reaching a contradiction.
Similarly, we can show $|{\cal X}'_{{\cal J}_j}| \leq k+1$.

\vspace{5pt}
\noindent
{\it Case 3:} $SP_j$ is computed by $GreedySP(S, i)$ in Step 3.
$SP_j$ is the leftmost stacking pair when it is chosen.
Let $SP_j = (s_p, s_{p+1}; s_{q-1}, s_q)$.
By the same approach as in Case 2,
we can show $|{\cal X}'_{{\cal I}_j}|, |{\cal X}'_{{\cal J}_j}| \leq 2$.
We further claim $|{\cal X}'_{{\cal I}_j}| \leq 1$.
Then $|SP_j| / | {\cal X}'_{{\cal I}_j} \cup {\cal X}'_{{\cal J}_j}| \geq 1/(1+2) = 1/3$.

To verify $|{\cal X}'_{{\cal I}_j}| \leq 1$,
we consider all possible cases with $|{\cal X}'_{{\cal I}_j}| = 2$
while there are no two consecutive stacking pairs.
The only possible case is that for some integers $r, t$,
both $(s_{p-1}, s_p; s_{r-1}, s_r)$
and $(s_p, s_{p+1}; s_{t-1}, s_t)$ belong to ${\cal X}'_{{\cal I}_j}$.
Then, $SP_j$ cannot be the leftmost stacking pair formed by unmarked bases,
contradicting the selection criteria of $SP_j$.
\end{proof}

\begin{theorem}
Let $S$ be an RNA sequence. Let $N^*$ be the maximum number of stacking
pairs that can be formed by any secondary structure of $S$. Let
$N$ be the number of stacking pairs output by $GreedySP(S,i)$. Then,
$N \geq \frac{N^*}{3}$.
\end{theorem}

\begin{proof}
By Lemmas \ref{sumfraction} and \ref{boundforspi}, the result follows.
\end{proof}

We remark that by setting $i=3$ in $GreedySP(S,i)$, we can already
achieve the approximation ratio of 1/3. The following theorem gives
the time and space complexity of the algorithm.

\begin{theorem}
Given an RNA sequence $S$ of length $n$ and a constant $k$, 
Algorithm \linebreak[4] $GreedySP(S,k)$
can be implemented in $O(n)$ time and $O(n)$ space.
\end{theorem}

\begin{proof}
Recall that the bases of an RNA sequence are chosen from the
alphabet $\{A,U,G,C\}$. If $k$ is a constant, there
are only constant number of different patterns of consecutive
stacking pairs that we must consider. For any $1 \leq j \leq k$,
there are only $4^j$ different strings that can be formed by
the four characters $\{A,U,G,C\}$. So, the locations of the
occurrences of these possible strings in the
RNA sequence can be recorded in an array of linked lists
indexed by the pattern of the string using $O(n)$ time preprocessing.
There are at most $4^j$ linked lists for any fixed $j$ and 
there are at most $n$ entries in these linked lists. In total,
there are at most $kn$ entries in all linked lists for all
possible values of $j$.

Now, we fix a constant $j$.
To locate all $j$ consecutive
stacking pairs,
we scan the RNA sequence from left to right. For each substring of
$j$ consecutive characters, we look up the array to see whether
we can form $j$ consecutive stacking pairs. By simple
bookkeeping, we can keep track which bases have been used
already. Each entry in the linked lists will only be
scanned at most once, so 
the whole procedure takes only $O(n)$ time. Since $k$ is a constant,
we can repeat the whole procedure for $k$ different values of $j$, and the
total time complexity is still $O(n)$ time.
\end{proof}

\newcommand{\encode}[1]{\langle #1 \rangle}

\section{NP-completeness}

In this section, we show that it is NP-hard to find a planar
secondary structure with the largest number of stacking pairs.
We consider the following decision problem.
Given an RNA sequence $S$ and an integer $h$, we wish to determine
whether the largest possible number of stacking pairs in a planar
secondary structure of $S$, denoted sp($S$), is at least $h$.  Below we show
that this decision problem is NP-complete by reducing the tripartite
matching problem \cite{Garey:1979:CIG} to it, which is defined as follows.

Given three node sets $X$, $Y$, and $Z$ with the same cardinality
$n$ and
an edge set $E \subseteq X \times Y \times Z$ of size $m$,
the {\it tripartite matching problem} is to 
determine whether $E$ contains a perfect matching, i.e.,
a set of $n$ edges which touches every node of $X$, $Y$, and $Z$
exactly once.

The remainder of this section is organized as follows.
Section~\ref{sec-construction} shows how we construct in polynomial
time an RNA sequence $S_E$ and an integer $h$ from a given instance
$(X,Y,Z, E)$ of the tripartite matching problem, where $h$
depends on $n$ and $m$.  Section~\ref{sec-if} shows that if $E$
contains a perfect matching, then sp($S_E$) $\ge h$.
Section~\ref{sec-only-if} is the non-trivial part, showing that if $E$
does not contain a perfect matching, then sp($S_E$) $< h$.  Combining
these three sections, we can conclude that it is NP-hard to
maximize the
number of stacking pairs for planar RNA secondary structures.

\subsection{Construction of the RNA sequence $S_E$} \label{sec-construction}

Consider any instance $(X,Y,Z, E)$ of the tripartite matching problem.
We construct an RNA sequence $S_E$ and an integer $h$ as follows.
Let $X = \{x_1, \cdots, x_n\}$, $Y = \{y_1, \cdots, y_n\}$, and $Z =
\{z_1, \cdots, z_n\}$.  Furthermore, let $E = \{ e_1, e_2, \cdots, e_m
\}$, where each edge $e_j = (x_{p_j}, y_{q_j}, z_{r_j})$.  Recall that
an RNA sequence contains characters chosen from the alphabet $\{A, U,
G, C\}$.  Below we denote $A^i$, where $i$ is any positive integer, as
the sequence of $i$ $A$'s. Furthermore, $A^+$ means a sequence of one
or more $A$'s.

\newcommand{\od}[1]{\overline{\delta(#1)}}
\newcommand{\op}[1]{\overline{\pi(#1)}}

Let $d = \max\{ 6n, 4(m+1) \} + 1$.  Define the following four RNA
sequences for every positive integer $k < d$.
\begin{itemize}
\item $\delta(k)$ is the sequence $U^dA^kGU^dA^{d-k}$, and
$\overline{\delta(k)}$ is the sequence $U^{d-k}A^dGU^kA^d$.
\item $\pi(k)$ is the sequence $C^{2d+2k} AG C^{4d-2k}$, and
$\overline{\pi(k)}$ is the sequence $G^{4d-2k}A G^{2d+2k}$.
\end{itemize}

{\small\bf Fragments:} Note that the sequences
$\delta(k)$ and $\od{k}$ are each composed of
two substrings in the form of $U^+ A^+$, separated by a character $G$.
Each of these two substrings is called a {\it fragment}. Similarly,
the two substrings of the form $C^+$ separated by $AG$ in $\pi(k)$
and the two substrings of the form $G^+$ separated by the character
$A$ in $\overline{\pi(k)}$ are also called fragments.

{\small\bf Node Encoding:} Each node in the three node sets $X$, $Y$,
and $Z$ is associated with a unique sequence.  For $1 \le i \le n$,
let $\encode{x_i}$, $\encode{y_i}$, $\encode{z_i}$ denote the
sequences $\delta(i)$, $\delta(n+i)$, $\delta(2n+i)$, respectively.
Intuitively, $\encode{x_i}$ is the encoding of the node $x_i$, and
similarly $\encode{y_i}$ and $\encode{z_i}$ are for the nodes $y_i$ and
$z_i$, respectively.  Furthermore, define $\encode{\overline{x_i}} =
\od{i}$, $\encode{\overline{y_i}} = \od{n+i}$, and
$\encode{\overline{z_i}} = \od{2n+i}$.

The node set $X$ is associated with two sequences $\cal X$ =
$\encode{x_1} G \encode{x_2} G \cdots G \encode{x_n}$ and
$\overline{\cal X}$ = $\encode{\overline{x_n}} G
\encode{\overline{x_{n-1}}} G \cdots G \encode{\overline{x_1}}$.
Let ${\cal X} - x_i$
= $\encode{x_1} G \cdots G \encode{x_{i-1}} G
\encode{x_{i+1}} G \cdots \encode{x_n}$ and $\overline{{\cal X} -
x_i}$ = $\encode{\overline{x_n}} G \cdots G
\encode{\overline{x_{i+1}}} G \encode{\overline{x_{i-1}}} G \cdots G
\encode{\overline{x_1}}$, where $x_i$ is any node in $X$.
Similarly, the node sets $Y$ and $Z$ are
associated with sequences ${\cal Y}$, $\overline{\cal Y}$, and
$\cal Z$, $\overline{\cal Z}$, respectively.

{\small\bf Edge Encoding:} For each edge $e_j$ (where $1 \le j \le
m$), we define four delimiter sequences, namely,
$V_j = \pi(j)$, $W_j = \pi(m+1+j)$, $\overline{V_j} = \overline{\pi(j)}$,
and $\overline{W_j} = \overline{\pi(m+1+j)}$.
Assume that $e_j = (x_{p_j}, y_{q_j},
z_{r_j})$. Then $e_j$ is encoded by the sequence $S_j$ defined as
\[ 
 AG~V_j~AG~W_j~AG~{\cal X}~G~{\cal Y}~G~{\cal Z}~G~
 \overline{({\cal Z} - z_{r_j})}~G~\overline{({\cal Y} - y_{q_j})}~G~
 \overline{({\cal X} - x_{p_j})}~\overline{V_j}~A~\overline{W_j}. 
\]
Let $S_{m+1}$ be a special sequence defined as $AG~V_{m+1}~AG~W_{m+1}~AG~
\overline{\cal Z}~G~\overline{\cal Y}~G~\overline{\cal X}~\;
\overline{V_{m+1}}~A~\overline{W_{m+1}}$.  In the following
discussion, each $S_j$ is referred to as a {\em region}.

Finally, we define $S_E$ to be the sequence $S_{m+1} S_m \cdots
S_1$.
Let $\sigma = 3n(3d-2) + 6d - 1$ and
let $h = m \sigma + n (6d - 4) + 12 d - 5$.  Note that $S_E$ has $O((n+m)^3)$
characters and can be constructed in $O(|S_E|)$ time.
In Sections \ref{sec-if} and \ref{sec-only-if}, we show that
sp($S_E$) $\ge h$ if and only if $E$ contains a perfect matching.

\subsection{Correctness of the if-part} \label{sec-if}
This section shows that if $E$ has a perfect matching,
we can construct a planar secondary structure for $S_E$
containing at least $h$ stacking pairs.  Therefore,
sp($S_E$) $\geq h$.

First of all, we establish several basic steps for constructing
stacking pairs on $S_E$.
\begin{itemize}
\setlength{\itemsep}{-1pt}
\item $\delta(i)$ or $\overline{\delta(i)}$ itself can form
      $d-1$ stacking pairs, while 
      $\delta(i)$ and $\overline{\delta(i)}$ together can form 
      $3d - 2$ stacking pairs.
\item 
        $\pi(i)$ and $\overline{\pi(i)}$ together can form 
        $6d - 2$ stacking pairs.
\item
For any $i \neq j$,
        $\pi(i)$ and $\overline{\pi(j)}$ together can form 
        $6d - 3$ stacking pairs.
\end{itemize}

\begin{lemma}
If $E$ has a perfect matching, then sp($S_E$) $\geq h$.
\end{lemma}
\begin{proof}
Let  $M = \{ e_{j_1}, e_{j_2}, \ldots, e_{j_n} \}$ be a perfect matching.
Without loss of generality, we assume that $1 \le j_1 < j_2 < \ldots < j_n
\le m$.  Define $j_{n+1} = m+1$.  
To obtain a planar secondary structure
for $S_E$ with at least $h$ stacking pairs, 
we consider the regions one by one. There are three cases.

\noindent
{\it Case 1:} We consider any region $S_j$ such that $e_j \not\in M$.
Our goal is to show that $\sigma = 3n(3d-2) +6d -1$ 
stacking pairs can be formed within $S_j$.  Note that
there are $(m-n)$ edges not in $M$.  Thus, we can obtain a total
of  $(m-n)\sigma$ stacking pairs in this case.  Details are as follows.
Assume that $e_j = (x_{p_j}, y_{q_j}, z_{r_j})$.
\begin{itemize}
\item $6d-2$ stacking pairs can be formed between $V_j$ and $\overline{V_j}$,
      and between $W_j$ and $\overline{W_j}$.
\item $3d-2$ stacking pairs can be formed 
        between $\encode{x_i}$ and $\encode{\overline{x_i}}$ 
        for all $i \neq p_j$, 
        and between $\encode{y_i}$ and $\encode{\overline{y_i}}$ 
        for all $i \neq q_j$,
        and between $\encode{z_i}$ and $\encode{\overline{z_i}}$ 
        for all $i \neq r_j$.
\item   $\encode{\overline{x_{p_j}}}$, $\encode{\overline{y_{q_j}}}$, and
        $\encode{\overline{z_{r_j}}}$ can each
        form $d-1$ stacking pairs.
\end{itemize}
The total  number of stacking pairs that can be formed within $S_j$
is $2(6d-2) + 3(n-1)(3d-2) + 3(d-1)$
= $3n(3d - 2) + 6d - 1$ = $\sigma$.

\noindent
{\it Case 2:} We consider the edges $e_{j_1}, e_{j_2}, \ldots, e_{j_n}$
in $M$. Our goal is to
show that each corresponding region  accounts for $\sigma + 6d -4$ 
stacking pairs. Thus, we obtain a total of $n\sigma + n(6d -4)$ stacking
pairs in this case.  Details are as follows.
Unlike Case 1, each region $S_{j_k}$, where $1 \le k \le n$,
may have some of its bases paired with that of $S_{j_{k+1}}$.
\begin{itemize}
\item $6d-3$ stacking pairs can be formed between $W_{j_k}$ in $S_{j_k}$
      and $\overline{W_{j_{k+1}}}$ in $S_{j_{k+1}}$.
\item $6d-2$ stacking pairs can be formed between $V_{j_k}$ in $S_{j_k}$
      and $\overline{V_{j_k}}$ in $S_{j_k}$.

\item $3d-2$ stacking pairs can be paired between $\encode{x_i}$ in
      $S_{j_k}$
      and $\encode{\overline{x_i}}$ in $S_{j_k}$ for any
      $i \neq p_{j_1}, \ldots, p_{j_k}$,
      and between $\encode{y_i}$ in $S_{j_k}$
   and $\encode{\overline{y_i}}$ in $S_{j_k}$ for any
   $i \neq q_{j_1}, \ldots, q_{j_k}$, and
   between $\encode{z_i}$ in $S_{j_k}$
   and $\encode{\overline{z_i}}$ in $S_{j_k}$ for any $i \neq r_{j_1}, 
   \ldots, r_{j_k}$.

\item
      $3d-2$ stacking pairs can be paired between $\encode{x_i}$ in
      $S_{j_k}$  and $\encode{\overline{x_i}}$ in $S_{j_{k+1}}$ for any
      $i = p_{j_1}, \ldots, p_{j_k}$,
      and between $\encode{y_i}$ in $S_{j_k}$
      and $\encode{\overline{y_i}}$ in $S_{j_{k+1}}$ for any
      $i = q_{j_1}, \ldots, q_{j_k}$, and
      between $\encode{z_i}$ in $S_{j_{k+1}}$
      and $\encode{\overline{z_i}}$ in $S_{j_{k+1}}$ for any $i = r_{j_1}, 
     \ldots, r_{j_k}$.

\end{itemize}
The total number of stacking pairs charged to $S_{j_k}$ is
$6d-3 + 6d -2 + 3n (3d -2)$ = $\sigma + 6d - 4$.  

\noindent
{\it Case 3:} We consider $S_{m+1}$.
We can form $6d-2$ stacking pairs between $V_{m+1}$ and 
$\overline{V_{m+1}}$, and
$6d-3$ stacking pairs between $W_{m+1}$ and $\overline{W_{j_1}}$.
The number of such stacking pairs is $12d - 5$.

Combining the three cases, the number of stacking pairs that
can be formed on $S_E$ is $(m-n)\sigma + n(\sigma + 6d - 4) + 12d - 5$,
which is exactly $h$.  Notice that no two stacking pairs formed
cross each other.  Thus, sp($S_E$) $\ge h$.
\end{proof}

\subsection{Correctness of the only-if part} \label{sec-only-if}

This section shows that if $E$ has no perfect matching, then
sp($S_E$)$<h$. We first give the framework of the proof in
Section~\ref{sec-only-if-framework}. 
Then, some basic definitions and concepts are
presented in Section~\ref{sec-only-if-definition}.
The proof of the only-if part
is given in Section~\ref{sec-only-if-proof}.

\newcommand{\opt}{\mbox{\rm OPT}}

\subsubsection{Framework of the proof} \label{sec-only-if-framework}
Let $\opt$ be a secondary structure of $S_E$ with the maximum
number of stacking pairs. Let $\#\opt$ be the number of stacking pairs
in $\opt$. That is, $\#\opt =$ sp($S_E$). In this section,
we will establish
an upper bound for $\#\opt$. Recall that we only consider
Watson-Crick base pairs, i.e., $A-U$ and $C-G$ pairs.
We define a conjugate of a
substring in $S_E$ as follows.

\vspace{5pt}
\noindent
{\bf Conjugates:}
For every substring $R = s_1 s_2 \ldots s_k$ of $S_E$,
the {\it conjugate} of $R$ is 
$\hat{R} = \hat{s_k} \ldots \hat{s_1}$,
where $\hat{A} = U$, $\hat{U} = A$, $\hat{C} = G$, and $\hat{G} = C$.

\vspace{5pt}
For example, $AA$'s conjugate is $UU$ and $UA$'s conjugate is $UA$.
To form a stacking pair, two adjacent bases must be paired
with another two adjacent bases. So, we concentrate on the possible
patterns of adjacent bases in $S_E$.

\vspace{5pt}
\noindent
{\bf 2-substrings:}
In $S_E$, any two adjacent characters are referred to as a 2-substring.
By construction, $S_E$ has only ten different types of 2-substrings:
$UU$, $AA$, $UA$, $GG$, $CC$, $GC$, $AG$, $GA$, $GU$, and $CA$-substrings.
A 2-substring can only form a stacking pair with its conjugate.
If they actually form a stacking pair in $OPT$, they are said to
be {\it paired}.

\vspace{5pt}
Since the conjugates of $AG$, $GA$, $GU$, and $CA$-substrings do not
exist in $S_E$, 
there is no stacking pair in $S_E$ which involves these 2-substrings.
We only need to consider $AA$, $UU$, $UA$, $GG$, $CC$, $GC$-substrings.
Table \ref{occ-2substrings} shows the numbers of
occurrences of these 2-substrings
in $S_j$ ($1 \leq j \le m+1$) and the total occurrences of these
substrings in $S_E$.

{\begin{table*}
\footnotesize
\begin{center}
\begin{tabular}{|l|l|l||l|}
\hline
Substring & \multicolumn{3}{c|}{Total number of occurrences of $t$ in} \\ \cline{2-4}
($t$) & $S_j$ ($j=1,2, \ldots, m$) & $S_{m+1}$ & $S_E$ \\ \hline
AA & $3n(d-2)+(3n-3)(2d-2)$ & $3n(2d-2)$ & $m(3n(d-2) + (3n-3)(2d-2)) + 3n(2d-2)$\\
UU & $3n(2d-2)+(3n-3)(d-2)$ & $3n(d-2)$ & $m(3n(2d-2) + (3n-3)(d-2)) + 3n(d-2)$\\
UA & $2(6n-3)$ & $6n$ & $2m(6n-3) + 6n$\\
GG & $2(6d-2)$ & $2(6d-2)$ & $2(m+1)(6d-2)$\\
CC & $2(6d-2)$ & $2(6d-2)$ & $2(m+1)(6d-2)$\\
GC & $4$ & $4$ & $4m+4$\\ \hline
\end{tabular}
\caption{Number of occurrences of different 2-substrings}
\label{occ-2substrings}
\end{center}
\end{table*}
}

Let $\#AA$ denote the number of occurrences of $AA$-substrings in $S_E$.
We use the $\#$ notation for other types of 2-subtrings in $S_E$ similarly.
The following fact gives a straightforward upper bound for $\#\opt$.

\begin{fact}  \label{lem-interval-very-basic}
\begin{tabbing}
ABCDEF \= $\#\opt$ \= $\le$ \= \kill
\> $\#\opt$ \> $\le$ \> $\min\{\#AA, \#UU\} + \min\{\#GG, \#CC\} + \#UA / 2 + \#GC / 2$ \\
\> \> $=$ \> $h + n + 1 + (2m+2)$.
\end{tabbing}
\end{fact}

Note that $\opt$ may not pair all $AA$-subtrings with $UU$-substrings.
Let $\diamondsuit AA$ be the number of $AA$-substrings that
are not paired in $\opt$.  Again, we use the $\diamondsuit$ notaion
for other types of 2-substrings. 
Fact~\ref{lem-interval-very-basic} can be strengthened as follows.

\begin{fact} \label{lem-interval-basic}
$\#\opt \le \min\{\#AA-\diamondsuit AA, \#UU-\diamondsuit UU\} + 
\min\{\#GG-\diamondsuit GG, \#CC-\diamondsuit CC\} +
(\#UA-\diamondsuit UA)/2 + (\#GC-\diamondsuit GC) / 2$.
\end{fact}

The upper bound given in Fact \ref{lem-interval-basic} forms
the basis of our proof for showing that $\#\opt < h$.
In the following sections, we consider the possible structure of
$\opt$. For each possible case, we show that the lower
bounds for some $\diamondsuit$ values, such as 
$\diamondsuit AA$ and $\diamondsuit CC$, are sufficiently
large so that $\opt$ can be shown to be less than $h$.
In particular, in one of the cases, we must make use of the fact
that $E$ does not have a perfect matching in order to prove the
lower bound for $\diamondsuit AA$, $\diamondsuit UA$, and $\diamondsuit
UU$. We give some basic definitions and concepts in Section
\ref{sec-only-if-definition}. The lower bounds and the
proof are given in Section \ref{sec-only-if-proof}.

\subsubsection{Definitions and concepts} \label{sec-only-if-definition}
In this section, we give some definitions and concepts which are
useful in deriving lower bounds for $\diamondsuit$ values.
We first classify each region $S_j$ in $S_E$
as either {\it open} or {\it closed} with
respect to $\opt$. Then, extending the definitions of fragments and
conjugates, we introduce {\it conjugate fragments} and
{\it delimiter fragments}. Finally, we present a property
of delimiter fragments in open regions.

\paragraph{Open and closed regions:}
With respect to $\opt$, a region
$S_j$ in $S_E$ is said to be an {\it open region}
if some $UU$, $AA$, or $UA$-substrings in $S_j$ are paired 
with some 2-substrings outside $S_j$;
otherwise, it is a {\it closed region}.

\begin{lemma} \label{lem-s_m+1}
If $S_{m+1}$ is a closed region, then $\#\opt < h$.
\end{lemma}
\begin{proof}
$S_{m+1}$ has $3nd$ more $AA$-substrings than $UU$-substrings.
If $S_{m+1}$ is a closed region, these $3nd$ $AA$-substrings
are not paired by $\opt$.
Thus, $\diamondsuit AA \geq 3nd$.
By Fact~\ref{lem-interval-basic}, $\#\opt < h+(n+1) + (2m+2) - 3nd < h$.
\end{proof}

Recall that $S_E$ is a sequence
composed of $\delta$'s, $\overline{\delta}$'s,
$\pi$'s, and $\overline{\pi}$'s.
Each $\delta(k)$ (respectively $\overline{\delta(k)}$) consists of 
two substrings of the form $U^+ A^+$, each of these substrings
is called a {\em fragment}.  Furthermore,
each $\pi(k)$ (resp.\ $\overline{\pi(k)}$) consists of
two substrings of the form $C^+$ (respectively $G^+$), each of these
subtrings is also called a fragment.

\paragraph{Conjugate fragments and delimiter fragments:}
Consider any fragment $F$ in $S_E$.
Another fragment $F'$ in $S_E$ is called a {\em conjugate fragment}\/
of $F$ if $F'$ is the conjugate of $F$.
Note that if $F$ is a fragment of a certian $\delta(k)$ (resp. $\pi(k)$), then
$F'$ appears only in some $\overline{\delta(k)}$ (respectively
$\overline{\pi(k)}$),
and vice versa.
By construction, if $F$ is a fragment of some delimiter sequence 
$V_j$ or $W_j$, then
$F$ has a unique conjugate fragment in $S_E$, which
is located in $\overline{V_j}$ or $\overline{W_j}$, respectively.
However, if $F$ is a fragment of some non-delimiter sequence,
says, $\encode{x_i}$, then for every instance of $\encode{\overline{x_i}}$ in $S_E$,
$F$ contains one conjugate fragment in $\encode{\overline{x_i}}$.

A fragment $F$ is said to be {\em paired}\/ with
its conjugate fragment $F'$ by $\opt$ if $\opt$ includes
all the pairs of bases between $F$ and $F'$.

For $1 \leq j \leq m+1$,
the fragment $F$ in $V_j$ or $W_j$
is called a {\it delimiter fragment}.
Note that the delimiter fragment $F$ should be of
the form $C^{2d+k}$ for $2d > k > 0$.

The following lemma shows a property of delimiter fragments
in open regions.

\begin{lemma} \label{lem-delimiter-fragment}
If $S_j$ is an open region, then both delimiter
fragments of either $V_j$ or $W_j$
must not pair with their conjugate fragments in $\opt$.
\end{lemma}
\begin{proof}
We prove the statement by contradiction.
Suppose one fragment of $V_j$ and one fragment of $W_j$
are paired with their conjugate fragments. 
Let $(s_x, s_{x+1}; s_{y-1}, s_y)$ and $(s_{x'}, s_{x'+1}; s_{y'-1}, s_{y'})$
be some particular stacking pairs in $V_j$ and $W_j$, respectively.
Since $S_j$ is an open region,
we can identify a stacking pair $(s_{x''}, s_{x''+1}; s_{y''-1}, s_{y''})$
where $s_{x''} s_{x''+1}$ and $s_{y''-1} s_{y''}$
are 2-substrings within and outside $S_j$, respectively.
Note that these three stacking pairs form an interleaving block.
By Lemma~\ref{interleavingblock}, ${\opt}$ is not planar,
reaching a contradiction.
\end{proof}

\subsubsection{Proof of the only-if part} \label{sec-only-if-proof}
By Lemma~\ref{lem-s_m+1}, it suffices to assume that
$S_{m+1}$ is an open region.
Before we give the proof of the only-if part, let us consider the
following lemma.

\begin{lemma} \label{lem-open-delimiter}
Let $\alpha$ be the number of delimiter fragments that
are not paired with their conjugate fragments.
Then,
$\diamondsuit CC + \diamondsuit GG \geq \alpha + (\#GC - \diamondsuit GC)$.
\end{lemma}
\begin{proof}
By construction, a $GC$-substring
must be next to the left end of a delimiter fragment $F$, which is
of the form $C^+$.
No other $GC$-substrings can exist. If this $GC$-substring is
paired, the leftmost $CC$-substring of $F$
must not be paired as there is no $GGC$ pattern in $S_E$.
Thus, $F$ must be one of the $\alpha$ delimiter fragments
that are not paired with their conjugate fragments.
Based on this observation, we classify
the $\alpha$ delimiter fragments into two groups:
(1) $(\#GC - \diamondsuit GC)$'s delimiter fragments whose
$GC$-substrings at the left end are paired; and
(2) $\alpha - (\#GC - \diamondsuit GC)$'s delimiter fragments whose
$GC$-substrings at the left end are not paired.

For each delimiter fragment $F = C^{2d+k}$ in group (1),
since the $GC$-substring on the left of $F$ is paired,
the leftmost $CC$-substring of $F$ must not be paired by $\opt$.
For the remaining $2d+k-2$ $CC$-substrings,
we either find a $CC$-substring which is not paired by $\opt$;
or these $2d+k-2$ $CC$-substrings are paired to
$GG$-substrings in some fragment $F' = G^{2d+k'}$ with $2d > k' > k$,
and thus, some $GG$-substring of $F'$ is not paired.
Therefore, each delimiter fragment in group (1) introduces
either (i) two unpaired $CC$-substrings or
(ii) one unpaired $CC$-substring and one unpaired $GG$-substring.
Hence, the total number of unpaired $CC$ and $GG$-substrings due to
delimiter fragments in group (1) $\geq 2 (\#GC - \diamondsuit GC)$.

For each delimiter fragment $F = C^{2d+k}$ in group (2), consider
the $CC$-substrings in $F$. With a similar argument, we can show
that 
each delimiter fragment in group (2) introduces
either (i) one unpaired $CC$-substring
or (ii) one unpaired $GG$-substring.
Hence, the total number of unpaired $CC$ and $GG$-substrings due to 
delimiter fragments in group (2) $\geq \alpha - (\#GC - \diamondsuit GC)$.

In total, we have
$\diamondsuit CC + \diamondsuit GG 
\geq \alpha + (\# GC - \diamondsuit GC)$.
\end{proof}

Now, we state a lemma which shows the
lower bounds for some $\diamondsuit$ values in terms of
the number of open regions in $\opt$.

\begin{lemma} \label{diamondlowerbounds}
Let $\ell \ge 1$ be the number of open regions in $\opt$.

\vspace{3pt}
\noindent
(1) If $S_{m+1}$ is an open region, then $\diamondsuit UU \geq 3(m+1-\ell) d$.

\vspace{3pt}
\noindent
(2) $\max \{ \diamondsuit CC, \diamondsuit GG \} \geq
    \ell + (\# GC - \diamondsuit GC) / 2$.

\vspace{3pt}
\noindent
(3) If $\ell = n+1$, $S_{m+1}$ is an open region,
    and $E$ does not have a perfect matching,
    then either (a) $\diamondsuit UU \geq 3(m-n)d + 1$,
    (b) $\diamondsuit AA \geq 1$, or (c) $\diamondsuit UA \geq 2$.
\end{lemma}

\begin{proof}

\noindent
{\small \bf Statement 1.}
Within each closed region $S_j$ where $j \neq m+1$,
$3d$'s $UU$-substrings cannot paired in $\opt$.
As there are $m+1-\ell$ such closed regions, $3(m+1-\ell)d$
$UU$-substrings are not
paired in $\opt$. Thus, $\diamondsuit UU \geq 3(m+1-\ell)d$.

\vspace{5pt}
\noindent
{\small \bf Statement 2.}
By Lemma~\ref{lem-delimiter-fragment}, we can identify $2 \ell$ fragments
in $V_j$ and $W_j$ of all open regions 
which are not paired with their conjugate fragments.
Then, by Lemma \ref{lem-open-delimiter}, we have
$\diamondsuit CC + \diamondsuit GG \geq 2\ell + (\# GC - \diamondsuit GC)$.
Thus, $\max\{ \diamondsuit CC, \diamondsuit GG \} \geq 
\ell + (\# GC - \diamondsuit GC) / 2$.

\vspace{5pt}
\noindent
{\small \bf Statement 3.}
By a similar argument to the proof for Statement 1, 
within the $m+1-\ell = m-n$ closed regions,
$3(m-n)d$ $UU$-substrings are not paired in $\opt$.

For the $\ell = n+1$ open regions,
one of them must be $S_{m+1}$. 
Let 
$S_{j_1}, \ldots, S_{j_n}$ be the remaining $n$ open regions.
Recall that $e_{j_1}, \ldots, e_{j_n}$
are the corresponding edges of these $n$ open regions.
Since these $n$ edges cannot form a perfect matching,
some node, says $x_k$, is adjacent to these $n$
edges more than once.
Thus, within $S_{j_1}, \ldots, S_{j_n}, S_{m+1}$,
we have more $\encode{x_k}$ than
$\encode{\overline{x_k}}$.
Therefore, at least two of the fragments in all $\encode{x_k}$
are not paired 
with their conjugate fragments.

Let $F$ be one of such fragments.
Note that $F$ is of the form $U^d A^k$.
Since $F$ is not paired with its conjugate fragment,
one of the following three cases occurs in $\opt$:

\vspace{3pt}
\noindent
Case 1: An $UU$-substring of $F$ is not paired.

\vspace{3pt}
\noindent
Case 2: An $AA$-substring of $F$ is not paired.

\vspace{3pt}
\noindent
Case 3: All $UU$-substrings and $AA$-substrings $F$ are paired.
In this case, $U^d$ of $F$ is paired with $A^d$ of a fragment
$F' = U^{k'}A^d$;
and $A^k$ of $F$ is paired with some substring $U^k$ of some fragment $F''$.
As $F'$ and $F''$ are not the same fragment, the $UA$-substrings of both $F$
and $F'$ are not paired.

\vspace{3pt}
In summary, we have either
(1) $\diamondsuit UU \geq 3(m-n)d + 1$, or
(2) $\diamondsuit AA \geq 1$, or
(3) $\diamondsuit UA \geq 2$.
\end{proof}

Based on Lemma \ref{diamondlowerbounds}, we prove the only-if part
by a case analysis in the following lemma.

\begin{lemma}
If $E$ does not have a prefect matching,
then $\# \opt < h$.
\end{lemma}
\begin{proof}
Recall that if $S_{m+1}$ is a closed region, then
$\#\opt < h$. Now, suppose that $S_{m+1}$ is an
open region. We show
$\# \opt < h$ in three cases $\ell < n+1$, $\ell > n+1$ and $\ell = n+1$.

\vspace{5pt}
\noindent
{\it Case 1:} $\ell < n+1$. By Lemma ~\ref{diamondlowerbounds} (1),
$\diamondsuit UU \geq 3(m+1-\ell)d$.
By Fact~\ref{lem-interval-basic},
we can conclude that $\#\opt = h + n+1 + (2m+2) - 3(n+1-\ell)d
\leq h + n+1 + (2m+2) - 3d < h$.

\vspace{5pt}
\noindent
{\it Case 2:} $\ell > n+1$. By Lemma~\ref{diamondlowerbounds} (2), 
$\max \{ \diamondsuit CC, \diamondsuit GG \} \geq \ell + (\# GC - \diamondsuit GC)/2$.
By Fact~\ref{lem-interval-basic},
$\#\opt \leq h + n + 1 - \ell$, which is smaller than $h$
because $\ell > n+1$.

\vspace{5pt}
\noindent
{\it Case 3:} $\ell = n+1$. By Lemma~\ref{diamondlowerbounds} (3),
either
(a) $\diamondsuit UU \geq 3(m-n)d + 1$, or
(b) $\diamondsuit AA \geq 1$, or
(c) $\diamondsuit UA \geq 2$.
By Fact~\ref{lem-interval-basic},
$\#\opt \leq h + n - \max \{ \diamondsuit CC, \diamondsuit GG \}
+ (\#GC - \diamondsuit GC) / 2$.
By Lemma~\ref{diamondlowerbounds} (2),
we have $\#\opt < h$.
\end{proof}

We conclude that if $E$ does not have a prefect matching,
then $\#\opt < h$. Equivalently,
if $\#\opt \geq h$, then
$E$ has a prefect matching.

\section{Conclusions}
In this paper, we have studied the problem of predicting RNA secondary
structures that allow arbitrary pseudoknots with a simple free
energy function that is minimized when the number of stacking
pairs is maximized. We have proved that this problem is NP-hard if the
secondary structure is required to be planar. We conjecture that
the problem is also NP-hard for the general case.
We have also given two approximation algorithms for this problem with
worst-case approximation ratios of 1/2 and 1/3 for planar and general
secondary structures, respectively. It would be of interest to
improve these approximation ratios.

Another direction is to study the problem using
energy function that is minimized when the number of base pairs is
maximized. It is known that this problem can be solved in cubic time
if the secondary structure can be non-planar \cite{Nussinov:1978:ALM}.
However, the computational complexity of the problem is still open if the
secondary structure is required to be planar. We conjecture that
the problem becomes NP-hard under this additional condition.
We would like to point out that the observation that have
enabled us to visualize the planarity of stacking pairs on a rectangular
grid does not hold in case of maximizing base pairs.

\bibliographystyle{plain}
\bibliography{rnastruct}

\end{document}